# Applications of the Characteristic Theory to the Madelung-de Broglie-Bohm System of Partial Differential Equations: The Guiding Equation as the Characteristic Velocity

Javier González,(1) Xavier Giménez,(2) Josep Maria Bofill(3)

(1) iBiTec-S, CEA-Saclay; javier-jose.gonzalez-aguilar@cea.fr
(2) Departament de Química Física and Institut de Química Teòrica i Computacional de la Universitat de Barcelona, Universitat de Barcelona; xgimenez@ub.edu
(3) Departament de Química Orgànica and Institut de Química Teòrica i Computacional de la Universitat de Barcelona, Universitat de Barcelona; jmbofill@ub.edu

**Abstract**

First, we use the theory of characteristics of first order partial differential equations to derive the guiding equation directly from the Quantum Evolution Equation (QEE). After obtaining the general result, we apply it to a set of evolution equations (Schrödinger, Pauli, Klein-Gordon, Dirac) to show how the guiding equation is, actually, the characteristic velocity of the corresponding matter field equations.

## I. Introduction

The quantum phenomena every time demands new point of view of the nature and their no classical concepts. In their explanation, Heisenberg's Matrix Mechanics and Schrödinger's Wave Mechanics gained quickly the favor of the community, leaving the classical concept of trajectory out of the standard of calculations.

Notwithstanding, through the independent work of Madelung,[1] de Broglie[2] and Bohm,[3] a new view of Quantum Mechanics was proposed which included the concept of trajectory, in a way analog to the classical counterpart. The formulation was at the beginning only for the physics explained by Schrödinger's equation,[1,3] but soon theories for Klein-Gordon,[2] Pauli,[4] Dirac,[5] Quantum Field Theory[6] and even the fundamental particles[7] appeared).

The idea of using trajectories for explaining quantum phenomena was firstly received with interest, although it didn't spread widely due to some features like the difficulty of computations, the fact that it did seem to add nothing to Wave Mechanics and its attachment to the hidden variables interpretation of quantum mechanics.

In despite of these problems, this formulation, called Hydrodynamic Formulation or Bohmian Mechanics, can be found continuously in the literature, with some important names like Takabayasi,[8] Vigier,[9] Holland,[10] etc. Even more, the practical use of Bohmian Mechanics have been increasing, especially in the fields of condensed matter,[11] chemical physics,[12] and in the study of time dependent systems.[13]

The basic idea under the use if trajectories in quantum mechanics is quite simple, both mathematically and physically, if we think in the position representation (it is also possible to use momentum representation, but it is more "natural" to use positions to find trajectories). In position representation, the wave function depends on position **x** and time *t*, so it is formally a valid question to ask how positions change in time, so which are their trajectories $\mathbf{x}(t)$.

This rate of change, the trajectory, plays a fundamental role in this formulation. In this article we use the mathematical theory of characteristics to solve the generalized Cauchy problem related to first order partial differential equations[14,15] (FO-PDE) to derive in a natural way the guiding equation as the characteristic velocity of the Madelung-de Broglie-Bohm (MdBB) equations, the matter field equations resulting from Schrödinger's equation. We also generalize this result to a variety of Quantum Evolution Equations, once their corresponding matter field equations have been derived.

It is important to note that the usual way to introduce the guiding equation, as explained below, is an analogy with Fluid Mechanics in which the current velocity of the "fluid" is assigned to the trajectories. In the derivation here presented, however, we recognize the guiding equation as the characteristic velocity of a system of equations, so we can establish it as a mathematical property of the equations, and not just as a physical analogy that needs to be added as an extra postulate to the theory.[10]

First of all, we derive briefly the mathematical expressions of the MdBB equations. From a mathematical point of view, they are defined by the set of equations,[3]

$$\frac{\partial \left[R(\mathbf{x},t)^2\right]}{\partial t} + \nabla \cdot \left[\frac{R(\mathbf{x},t)^2 \nabla S(\mathbf{x},t)}{m}\right] = 0 \tag{1}$$

$$\frac{\partial S(\mathbf{x},t)}{\partial t} + \frac{[\nabla S(\mathbf{x},t)]^2}{2m} + V(\mathbf{x},t) - \frac{\hbar^2}{2m} \frac{\nabla^2 R(\mathbf{x},t)}{R(\mathbf{x},t)} = 0 \tag{2}$$

$$\frac{d\mathbf{x}}{dt} = \mathbf{v}(\mathbf{x},t) = \frac{\nabla S(\mathbf{x},t)}{m} \tag{3}$$

In equations (1)-(3), $R(\mathbf{x},t)$ and $S(\mathbf{x},t)$ are real functions of the real variables **x** (the space vector) and *t* (the time); *m* is the mass of the particle and $\hbar = h/2\pi$, with *h* being Planck's constant.

Equations (1) and (2) are called the Madelung-de Broglie-Bohm (MdBB) equations. They are the result of writing Schrödinger's equation with the wave function in the polar form $\Psi(\mathbf{x},t) = R(\mathbf{x},t) \exp[iS(\mathbf{x},t)/\hbar]$. Although they seem to be singular in nodes, i.e. where $R(\mathbf{x},t) = 0$ -the points where the particle cannot be-, it is just an artifact due to the way in which equations (1)-(3) are written.

Following the usual presentation,[10,16] equation (1) is called the continuity equation, since the definition of the quantum flux as

$$\mathbf{j}(\mathbf{x},t) = R(\mathbf{x},t)^2 \nabla S(\mathbf{x},t)/m \qquad (4)$$

Allows writing equation (1) as

$$\frac{\partial \left[R(\mathbf{x},t)^2\right]}{\partial t} + \nabla \cdot \left[\mathbf{j}(\mathbf{x},t)\right] = 0$$

This is a continuity equation for the quantum flux $\mathbf{j}$. It is responsible, then, of the evolution of $R(\mathbf{x},t)$, the modulus of the wave function.

Equation (2) deals with the evolution of $S(\mathbf{x},t)$, the phase of the wave function. It looks like the classical Hamilton-Jacobi equation with an extra term $Q(\mathbf{x},t) = -\frac{\hbar^2}{2m} \frac{\nabla^2 R(\mathbf{x},t)}{R(\mathbf{x},t)}$, called the quantum potential. We define then the total potential,

$$W(\mathbf{x},t) = V(\mathbf{x},t) - \frac{\hbar^2}{2m} \frac{\nabla^2 R(\mathbf{x},t)}{R(\mathbf{x},t)} \qquad (5)$$

This will be very useful to study the quantum evolution of the system, as has been shown in a transmission context.[17]

Finally, equation (3) is the so-called guiding equation. It allows defining quantum trajectories by giving the evolution of their tangents. Equation (3) has been justified in several ways, from analogies,[2,3] statistics[18] and dynamics.[19] Perhaps, the easiest way is the analogy of the quantum flux. Since a flux can be written as $\mathbf{j}(\mathbf{x},t) = R(\mathbf{x},t)^2 \mathbf{v}(\mathbf{x},t)$, comparison with the definition (4) gives the guiding equation (3) as the current velocity. Notwithstanding, there is no prove that the guiding equation is the equation of the trajectories $\mathbf{x}(t)$, so it has to be added to the theory as an extra postulate, as Holland did in his book.[10]

However, we want to show that the system of equations (1)-(3) is redundant. In particular, that equation (3) is already included in equations (1)-(2). Even more, since equations (1)-(2) are the Schrödinger equation with the wave function in polar form, that means that trajectories exist in Quantum Mechanics naturally, with no need to add an extra postulate. Indeed, although our argumentation deals with the Schrödinger equation, we will show that it can be extended to all quantum evolution equations.

For our purpose, in the next section, we will derive the guiding equation (3) using the method of characteristics; in fact, we will realize that the guiding equation is one of the equations that define the characteristic curves of the MdBB equations (1)-(2). Section III will be devoted to the extension of the procedure to other quantum evolution equations, with the aim to show the generality of the approach. Finally, in section IV we address two important mathematical questions related to the theory, the total differential and the complete integral.

## II. The Characteristic Curves of MdBB Equations: Derivation of the Guiding Equation

First, let us remember here some of the results of the method of characteristics in PDE, their proofs and explanation rest to the specific literature. Given a PDE for the unknown function $u(\mathbf{x},t)$,

$$F\left(t, \mathbf{x}, u, \frac{\partial u}{\partial t}, \nabla u\right) = 0 \tag{6}$$

Equation (6) can be related to an equivalent system of Ordinary Differential Equations (ODE) along the characteristic curves,

$$d t : d\mathbf{x} : d u : d\left(\frac{\partial u}{\partial t}\right) : d(\nabla u) =$$

$$= \frac{\partial F}{\partial(\partial u / \partial t)} : \nabla_{\nabla u} F : \nabla u \cdot \nabla_{\nabla u} F + \frac{\partial u}{\partial t}\frac{\partial F}{\partial(\partial u / \partial t)} : -\left(\frac{\partial u}{\partial t}\frac{\partial F}{\partial u} + \frac{\partial F}{\partial t}\right) : \nabla u \frac{\partial F}{\partial u} + \nabla F \tag{7}$$

In equation (7), the subindex of the nabla operator indicates with respect which variable derivation must be done, the appearance of no subindex meaning positions.

Equation (7) allows computing directly the guiding equation as the tangent of the characteristic curves,

$$\frac{d\mathbf{x}}{dt} = \frac{\nabla_{\nabla u} F}{\partial F / \partial(\partial u / \partial t)} \tag{8}$$

Although in equations (6)-(8) $(\mathbf{x},t)$ is a vector of generalized coordinates and a parameter that parameterize the curves, in the particular case when they are true spatial and temporal coordinates, equation (8) gives the tangent to the trajectories $\mathbf{x}(t)$, and then, the velocity. Equation (8) is, then, the basis of all our analysis, and we will refer to it continuously through this work.

Of course, equations (6)-(8) refer to a single equation with one unknown function. Nevertheless, their generalization to a system of equations is known,[14] and will be used if needed.

It could be a bit surprising the last part of the last statement. MdBB equations are a system of two equations with two unknowns, $R(\mathbf{x},t)$ and $S(\mathbf{x},t)$, so "if needed" seems to be "always". Notwithstanding, MdBB equations (1)-(2) can be also viewed as a set of two evolution equations; on one hand the continuity equation refers to the evolution of the density, the modulus squared of the wave function; and on the other hand the Hamilton-Jacobi equation refers to the evolution of the action, the phase of the wave function. In addition, then they are PDE of first order with respect to time.

Hence, we can treat them formally as a set of two equations,

$$G_1\left(R, \frac{\partial R}{\partial t}, \nabla R, \mathbf{x}, t\right) = 0$$
$$G_2\left(S, \frac{\partial S}{\partial t}, \nabla S, \mathbf{x}, t\right) = 0 \qquad (9)$$

In this way, the dependencies on $S(\mathbf{x},t)$ in the continuity equation and the dependencies on $R(\mathbf{x},t)$ in the Hamilton-Jacobi equation are ignored to be only dependencies on $\mathbf{x}$ and $t$. This means that the coupling between $R(\mathbf{x},t)$ and $S(\mathbf{x},t)$ is ignored in equations (9); the continuity equation, then, is a linear FO-PDE, whereas the Hamilton-Jacobi equation is a non-linear PDE. Although this approach could guide to confusion between the characteristic velocity and the flux velocity, in the cases under study both coincide if nothing is added on this subject.

Finally, we want to emphasize that it is the whole set of equations (7) which is equivalent, along the direction of the characteristics (8), to the original system of PDE (6). In spite of this fact, we will only write down all the system (7) in section IV, where we are concern with the completeness of the approach. The cause is that we are interested in the guiding equation, and sometimes in some other points related, but not in solving the system of equations (6). Therefore, we will write just the equations that we need for our developments, which normally will be

$$\frac{dR}{dt} = \frac{\nabla R \cdot \nabla_{\nabla R} G_1 + \frac{\partial R}{\partial t} \frac{\partial G_1}{\partial (\partial G_1 / \partial t)}}{\frac{\partial G_1}{\partial (\partial R / \partial t)}} \qquad (10)$$

$$\frac{d\nabla S}{dt} = \frac{\nabla S \frac{\partial G_2}{\partial S} + \nabla G_2}{\frac{\partial G_2}{\partial (\partial S / \partial t)}} \qquad (11)$$

Of course, equations (10)-(11) are valid only if the denominators are non-zero. To derive equations (10)-(11), equations (9) have been used and both equations must be understood as being valid along the characteristic curves which tangent is given by equation (8), i.e., they are valid along the trajectories. Equation (10) expresses the evolution of the modulus of the wave function, and is then closely related to the evolution of the density of the system under study. Equation (11) will reveal, see below, to be rise to the generalization along the trajectories of Newton's second law in a variety of cases.

Now we can begin with the study of Schrödinger's equation and the original MdBB equations, (1)-(2). Application of equation (8) to whatever of the equations (1)-(2) gives directly the guiding equation as,

$$\frac{\partial G_1}{\partial(\partial R/\partial t)} = \frac{\partial G_2}{\partial(\partial S/\partial t)} = 1 \qquad (12)$$

$$\frac{\partial G_1}{\nabla R} = \frac{\partial G_2}{\nabla S} = \frac{\nabla S}{m} \qquad (13)$$

Substitution in equation (8) gives directly the guiding equation,

$$\frac{d\mathbf{x}}{dt} = \frac{\nabla S(\mathbf{x},t)}{m} \qquad (14)$$

Equation (14) shows clearly that the guiding equation introduced by de Broglie and Bohm a bit arbitrarily is, in reality, a property inherent to the system of PDE form by the MdBB equations, since it gives the direction of their characteristic curves.

It is interesting to go a bit further in the use of the method of characteristics applied to MdBB equations. Along the lines defined by the trajectories, equations (10)-(11) give

$$\frac{dR(\mathbf{x},t)}{dt} = -\frac{R(\mathbf{x},t)}{2m}\nabla^2 S(\mathbf{x},t) \qquad (15)$$

$$\frac{d[\nabla S(\mathbf{x},t)]}{dt} = -\nabla W(\mathbf{x},t) \qquad (16)$$

Although equation (16) is formally easy to integrate to find the gradient of the action, it is very interesting to see that, using the guiding equation (14), equation (16) is equal to,

$$\frac{d^2\mathbf{x}}{dt^2} = -\frac{\nabla W(\mathbf{x},t)}{m} \qquad (17)$$

This, accepting $W(\mathbf{x},t)$ as the total potential that acts to the system, is the generalization of Newton's second law.

Although we have treated equations (1)-(2) as independent, finally, for completeness we want to show how to combine them to reduce the system of MdBB equations to a single integro-differential equation for the action. Equation (15) can be integrated,

$$R(\mathbf{x},t) = R(\mathbf{x},t_0)\exp\left(-\frac{1}{2m}\int_{t_0}^{t}\nabla^2 S(\mathbf{x}(t'),t')dt'\right) \qquad (18)$$

Where the integral has to be evaluated along the trajectory $\mathbf{x}(t)$. Substituting equation (18) in the total potential (5) and using the Quantum Hamilton-Jacobi equation (2) allows reducing Schrödinger's equation to the solution of the integro-differential equation,[17]

$$\frac{\partial S(\mathbf{x},t)}{\partial t} + \frac{[\nabla S(\mathbf{x},t)]^2}{2m} - \frac{\hbar^2}{2m}\left\{\frac{1}{4m^2}\left\{\int_{t_0}^{t}\nabla[\nabla^2 S(\mathbf{x}(t'),t')]dt'\right\}^2 - \right.$$
$$\left. -\frac{1}{2m}\int_{t_0}^{t}\nabla^4 S(\mathbf{x}(t'),t')dt'\right\} + V(\mathbf{x},t) = 0 \qquad (19)$$

It is important to note that because the integrals in equation (19) must be computed along the trajectory, it shows explicitly how the whole ensemble of quantum trajectories up to the desired time affects each quantum trajectory, i.e., the past evolution of the whole wave function is needed to compute the future evolution of each trajectory.

Finally, we want to remark a property of the action, presented as the phase of the wave function. When integrating the corresponding equation (7), one gets,

$$S(\mathbf{x},t) - S(\mathbf{x}_0,t_0) = \int_{t_0}^{t}\left[-V(\mathbf{x}(t'),t') + \frac{\hbar^2}{2m}\frac{\nabla^2 R(\mathbf{x}(t'),t')}{R(\mathbf{x}(t'),t')} - \frac{1}{2m}[\nabla S(\mathbf{x},t)]^2\right]dt' \qquad (20)$$

Where the definition of the total potential (5) has been used and the integral has to be evaluated along the trajectory $\mathbf{x}(t)$ with initial point, $\mathbf{x}_0 = \mathbf{x}(t_0)$. Equation (20) is analog to its classical counterpart for the action using a total potential as defined in (5). It is important to note that in equation (20), the function $S(\mathbf{x},t)$, is not uniquely fixed since any two functions, say, $S(\mathbf{x},t)$ and $S'(\mathbf{x},t)$, which differ by an integer multiple of $h$, given rise to the same wave function, $\Psi(\mathbf{x},t)$.

### III. Generalization to other quantum evolution equations

In this last section, we will derive the guiding equation of three different quantum evolution equations. Firstly, we will study some variations of Schrödinger equations, like Pauli equation and the case with the kinetic term position depending (which can be seen as a position-dependent mass or a curved geometry). Secondly, we will study other evolution equations like the Klein-Gordon equation or Weyl equation. Our purpose is not to be exhaustive, but to show that the derivation of the guiding equation from the quantum evolution equation with the aid of characteristic curves is general in Quantum Physics. Also we will show some of the interesting questions that this derivation rise when working with spinors, or when the kinetic operator depends on the first partial derivative, and not in the second one as in Schrödinger equation.

*III.1. Pauli equation*

The previous results seem to identify the velocity of Quantum Trajectories with internal properties of the quantum. Nevertheless, external fields can affect the velocity of Quantum Trajectories. An example is Pauli equation, which also will be very useful to show that the method of characteristics can be used to compute the velocity of Quantum Trajectories even when the Quantum Evolution Equation deals with spinors.

If $\Psi^T = (\psi_1,\psi_2)$ is a bicomponent spinor, Pauli equation, which approximates in a non-relativistic way the coupling of a ½-spin system to a magnetic field, reads,

$$i\hbar \frac{\partial}{\partial t} \mathbf{I}_2 \Psi(\mathbf{x},t) = \left[ \frac{1}{2m}(i\hbar\nabla - q\mathbf{A})^2 + q\phi \right] \mathbf{I}_2 \Psi(\mathbf{x},t) - \frac{q\hbar}{2m} \sigma \cdot (\nabla \times \mathbf{A}) \Psi(\mathbf{x},t) \quad (21)$$

Where $\sigma$ is the vector of Pauli matrices, $\sigma = (\sigma_1, \sigma_2, \sigma_3) = \left( \begin{pmatrix} 0 & 1 \\ 1 & 0 \end{pmatrix}, \begin{pmatrix} 0 & -i \\ i & 0 \end{pmatrix}, \begin{pmatrix} 1 & 0 \\ 0 & -1 \end{pmatrix} \right)$; $\mathbf{A}$ and $\phi$ are the electromagnetic potentials (vector and scalar, respectively) and $q$ is the charge.

The first point is to express the spinor in polar form. Two possibilities are common in the literature. On one hand, de Broglie introduced the expression,[20]

$$\psi_j(\mathbf{x},t) = R_j(\mathbf{x},t) \exp\left[ \frac{i}{\hbar} S_j(\mathbf{x},t) \right] \quad (22)$$

I.e., he wrote each component of the spinor in polar form. On the other hand, Holland uses[21]

$$\psi_j(\mathbf{x},t) = R(\mathbf{x},t) \exp\left[ \frac{i}{\hbar} S(\mathbf{x},t) \right] r_j(\mathbf{x},t) \exp\left[ \frac{i}{\hbar} s_j(\mathbf{x},t) \right] \quad (23)$$

I.e., he added two extra functions (also real functions of real variables) to account for the spin. As both formulations are easily related in this case, we will apply the method of characteristics as outlined in section II just to the first of them.

The MdBB equations for Pauli equation read,

$$\frac{\partial R_i(\mathbf{x},t)}{\partial t} + \nabla R_i(\mathbf{x},t) \cdot \left[ \frac{\nabla S_i(\mathbf{x},t)}{m} + \frac{q\mathbf{A}}{m} \right] + \frac{R_i(\mathbf{x},t) \nabla^2 S_i(\mathbf{x},t)}{2m} + C_i(\mathbf{x},t,R_i,S_i) = 0 \quad (24)$$

$$\frac{\partial S_i(\mathbf{x},t)}{\partial t} + \left[ \frac{\nabla S_i(\mathbf{x},t)}{2m} + \frac{q\mathbf{A}}{m} \right] \cdot \nabla S_i(\mathbf{x},t) + D_i(\mathbf{x},t,R_i,S_i) = 0 \quad (25)$$

Where $i = 1,2$ runs over the spinor indexes, and the functions $C$ and $D$ are obtained directly from the algebraic computation and contain the possible couplings between spinors and, $D$, also contains the quantum potential.

The fact that they are two equations for the modulus of the wave function, $G_1$ and $G_3$, and two equations for the actions, $G_2$ and $G_4$, does not require any kind of modification. Notwithstanding, the guiding equation is now double, with a different velocity for each spinor. Application of equation (8) gives the guiding equation,

$$\frac{d\mathbf{x}_1}{dt} = \frac{\partial G_1}{\nabla_{\nabla R_1} R_1} = \frac{\partial G_2}{\nabla_{\nabla S_1} S_1} = \frac{\nabla S_1(\mathbf{x},t)}{m} + \frac{q\mathbf{A}}{2m}$$
$$\frac{d\mathbf{x}_2}{dt} = \frac{\partial G_3}{\nabla_{\nabla R_2} R_2} = \frac{\partial G_4}{\nabla_{\nabla S_2} S_2} = \frac{\nabla S_2(\mathbf{x},t)}{m} + \frac{q\mathbf{A}}{2m}$$
(26)

Therefore, the well-known result of the guiding equation for the Pauli equation is recovered.

The coupling between spinors can be shown using equation (10) to compute the evolution of the modulus of the wave function. Its application to equations (24) gives for $R_1$ (we omit $R_2$ because it is almost the same),

$$\frac{dR_1(\mathbf{x},t)}{dt} = -R_1(\mathbf{x},t)\left[\frac{\nabla^2 S_1(\mathbf{x},t)}{2m} - \frac{q\nabla \cdot A}{m}\right] + \frac{qR_2(\mathbf{x},t)}{2m}\left[(\sin\xi,\cos\xi,0)\cdot \nabla \times \mathbf{A}\right]$$
(27)

with $\xi = \xi(\mathbf{x},t) = \dfrac{S_2(\mathbf{x},t) - S_1(\mathbf{x},t)}{\hbar}$.

Finally, the corresponding Newton's second law can be obtained applying equation (11) to equations (25). For the simplicity of the expressions, we limit ourselves to the case $\mathbf{A} = \mathbf{0}$, and obtain,

$$\frac{d^2\mathbf{x}_i}{dt^2} = -q\nabla\phi - \frac{\hbar^2}{2m}\frac{\nabla^2 R_i(\mathbf{x},t)}{R_i(\mathbf{x},t)}$$
(28)

*III.2. Non-constant mass Schrödinger equation*

The case of Schrödinger's equation when the kinetic part depends on the position and/or time can be called "non-constant" mass because it has the same effect of a non-constant mass. Recently, it has received an increasing interest because it appears in some emerging fields like quantum dots,[22] and chemical physics of transmission problems;[23] in addition, it appears also in problems of curvilinear coordinates,[24] path integral,[25] and Green function techniques.[26]

The evolution equation is

$$-\frac{\hbar^2}{2}\nabla\left[\frac{1}{m(\mathbf{x},t)}\nabla\Psi(\mathbf{x},t)\right] + V(\mathbf{x},t) = i\hbar\frac{\partial\Psi(\mathbf{x},t)}{\partial t}$$
(29)

Writing the wave function in polar form gives rise to two equations, one for the real and another for the imaginary part, which are the analog to MdBB equations,

$$\frac{\partial R(\mathbf{x},t)}{\partial t} + \nabla \cdot \left[\frac{R(\mathbf{x},t)\nabla S(\mathbf{x},t)}{m(\mathbf{x},t)}\right] = 0 \qquad (30)$$

$$\frac{\partial S(\mathbf{x},t)}{\partial t} + \frac{[\nabla S(\mathbf{x},t)]^2}{2m(\mathbf{x},t)} + V(\mathbf{x},t) - \frac{\hbar^2}{2m(\mathbf{x},t)} \frac{\nabla^2 R(\mathbf{x},t)}{R(\mathbf{x},t)} + \frac{\hbar^2 \nabla m(\mathbf{x},t) \cdot \nabla R(\mathbf{x},t)}{2m(\mathbf{x},t)^2 R(\mathbf{x},t)} = 0 \qquad (31)$$

Where we define now the total potential as,

$$\Omega(\mathbf{x},t) = V(\mathbf{x},t) - \frac{\hbar^2}{2m(\mathbf{x},t)} \frac{\nabla^2 R(\mathbf{x},t)}{R(\mathbf{x},t)} + \frac{\hbar^2 \nabla m(\mathbf{x},t) \cdot \nabla R(\mathbf{x},t)}{2m(\mathbf{x},t)^2 R(\mathbf{x},t)} \qquad (32)$$

Application of equation (8) to equations (30)-(31) gives,

$$\frac{d\mathbf{x}}{dt} = \frac{\partial G_1}{\nabla_{\nabla R} R} = \frac{\partial G}{\nabla_{\nabla S} S} = \frac{\nabla S(\mathbf{x},t)}{m(\mathbf{x},t)} \qquad (33)$$

Equation (33) is a result often used, also normally only justified because it is analog to the constant mass case, (14). This development, at contrary, allows obtaining it from the evolution equation.

As we did in section II for the Schrödinger's case, in order to know how the density varies, we apply equation (10),

$$\frac{dR(\mathbf{x},t)}{dt} = -\frac{R(\mathbf{x},t)}{2m(\mathbf{x},t)} \nabla^2 S(\mathbf{x},t) - \frac{R(\mathbf{x},t)\nabla S(\mathbf{x},t) \cdot \nabla m(\mathbf{x},t)}{m(\mathbf{x},t)^2} \qquad (34)$$

Equation (34) can be easily integrated to obtain the density,

$$R(\mathbf{x},t) = R(\mathbf{x},t_0)\exp\left(-\frac{1}{2}\int_{t_0}^{t}\left[\frac{\nabla^2 S(\mathbf{x}(t'),t')}{m(\mathbf{x}(t'),t')} + \frac{\nabla S(\mathbf{x}(t'),t') \cdot \nabla m(\mathbf{x}(t'),t')}{m(\mathbf{x}(t'),t')^2}\right]dt'\right) \qquad (35)$$

Equation (35) must be integrated along the trajectory. It shows how a non-constant mass affects the evolution of the density in a direct way.

Finally, application of equation (11) gives the generalization of Newton's second law,

$$\frac{d}{dt}\left[m(\mathbf{x},t)\frac{d\mathbf{x}}{dt}\right] = -\nabla\Omega(\mathbf{x},t) - \frac{1}{2}\left(\frac{d\mathbf{x}}{dt}\right)^2 \nabla m(\mathbf{x},t) \qquad (36)$$

Equation (36), as expected, is analog to the classical case with a total potential $\Omega(\mathbf{x},t)$. Finally, $S(\mathbf{x},t)$ must be integrated along the trajectory to obtain a result similar to that given in equation (20).

### III.3. Klein-Gordon equation

It was the case studied initially by de Broglie,[2] because it includes relativistic effects. For a free particle it reads,

$$\left[\Box^2 + \frac{m^2 c^2}{\hbar^2}\right]\Psi(\mathbf{x},t) = 0 \tag{37}$$

Where $c$ is the velocity of light and $\Box^2$ is the d'Alembertian operator, $\Box^2 = \frac{1}{c^2}\frac{\partial^2}{\partial t^2} - \nabla^2$. We follow the same scheme of previous sections. Hence, we begin with the polar form of $\Psi(\mathbf{x},t)$ to obtain the corresponding MdBB equations. After splitting in real and imaginary parts, we have,

$$\frac{\partial R(\mathbf{x},t)}{\partial t}\frac{\partial S(\mathbf{x},t)}{\partial t} - c^2 \nabla R(\mathbf{x},t) \cdot \nabla S(\mathbf{x},t) + R(\mathbf{x},t)\left[\frac{\partial^2 S(\mathbf{x},t)}{\partial t^2} - c^2 \nabla^2 S(\mathbf{x},t)\right] = 0 \tag{38}$$

$$\frac{1}{c^2}\left(\frac{\partial S(\mathbf{x},t)}{\partial t}\right)^2 - \frac{\hbar^2}{c^2 R(\mathbf{x},t)}\frac{\partial^2 R(\mathbf{x},t)}{\partial t^2} + \hbar^2 \frac{\nabla^2 R(\mathbf{x},t)}{R(\mathbf{x},t)} - [\nabla S(\mathbf{x},t)]^2 - m^2 c^2 = 0 \tag{39}$$

Or, in a more concise form,

$$\Box[R(\mathbf{x},t) \cdot \Box S(\mathbf{x},t)] = 0$$
$$R(\mathbf{x},t)[\Box S(\mathbf{x},t)]^2 - \hbar^2 \Box^2 R(\mathbf{x},t) = 0$$

So that the quantum potential can be defined as,

$$W(\mathbf{x},t) = \hbar^2 \frac{\Box^2 R(\mathbf{x},t)}{R(\mathbf{x},t)} \tag{40}$$

Equations (38)-(39) are the MdBB equations for the Klein-Gordon case. They are second order PDE even with respect to time. The approach given up to here must be, then, slightly modified. It is based in the characteristic curves of the MdBB equations. For equations (38)-(39), the characteristic direction coincides with that of the system

$$\frac{\partial \rho(\mathbf{x},t)}{\partial t}\frac{\partial \sigma(\mathbf{x},t)}{\partial t} - c^2 \nabla \rho(\mathbf{x},t) \cdot \nabla \sigma(\mathbf{x},t) + R(\mathbf{x},t)\left\{\left[\frac{\partial \sigma(\mathbf{x},t)}{\partial t}\right]^2 - c^2 [\nabla \sigma(\mathbf{x},t)]^2\right\} = 0 \tag{41}$$

$$\frac{1}{c^2}\left(\frac{\partial \sigma(\mathbf{x},t)}{\partial t}\right)^2 - \frac{\hbar^2}{c^2 \rho(\mathbf{x},t)}\left[\frac{\partial \rho(\mathbf{x},t)}{\partial t}\right] + \hbar^2 \frac{[\nabla \rho(\mathbf{x},t)]^2}{\rho(\mathbf{x},t)} - [\nabla \sigma(\mathbf{x},t)]^2 - m^2 c^2 = 0 \tag{42}$$

In equations (41)-(42), $\rho(\mathbf{x},t)$ and $\sigma(\mathbf{x},t)$ are the unknown functions. Now, following the same argumentation than in previous cases for equations (41)-(42), we apply equation (8) to obtain the guiding equation,

$$\frac{d\mathbf{x}}{dt} = \frac{\partial G_1/\nabla_{\nabla R} R}{\partial G_1/\partial(\partial R/\partial t)} = \frac{\partial G_2/\nabla_{\nabla S} S}{\partial G_2/\partial(\partial S/\partial t)} = -c^2 \frac{\nabla S(\mathbf{x},t)}{\partial S(\mathbf{x},t)/\partial t} \qquad (43)$$

Of course, this is result already proposed by de Broglie.[2] Interestingly; de Broglie worked with front waves, a subject closely related with the method of characteristics.

The density can be obtained if equation (10) is applied to equation (38),

$$\frac{dR(\mathbf{x},t)}{dt} = -R(\mathbf{x},t)\Box^2 S(\mathbf{x},t) \qquad (44)$$

The modulus of the wave function (and then the density), then can be obtained by integration of equation (44),

$$R(\mathbf{x},t) = R(\mathbf{x},t_0)\exp\left(-\int_{t_0}^{t}\Box^2 S(\mathbf{x}(t'),t')dt'\right) \qquad (45)$$

Where the integration must be done along the trajectory.

It is possible also to obtain a generalization of Newton's second law, since application of equation (11) to (38) gives, using the quantum potential (40),

$$\frac{d[\nabla S(\mathbf{x},t)]}{dt} = -\nabla W(\mathbf{x},t) \qquad (46)$$

Substitution of the guiding equation (43) and some arrangement gives,

$$\frac{d^2\mathbf{x}}{dt^2}\frac{\partial S(\mathbf{x},t)}{\partial t} = c^2 \frac{d\mathbf{x}}{dt}\cdot\Box W(\mathbf{x},t) \qquad (47)$$

As last note on this case, let us add that the guiding equation (43) comes up with a bizarre feature, since the velocity is not bounded by $c$. Moreover, from equations (38) or (39) we can define the probability density $\rho(\mathbf{x},t) = R(\mathbf{x},t)^2 \frac{\partial S(\mathbf{x},t)}{\partial t}$, which is not positive defined. These are well-known features of the Klein-Gordon equation and we won't discussed them because we just present the method, see the literature.[27] But we note them because recovering those features shows that with trajectories we have added nothing to the theory, due to the fact that trajectories are already implicit in it.

*III.4. Weyl equation (1D Dirac equation)*

We add this last example motivate by a recent result of Holland.[28] The 1D Dirac equation in Weyl representation, reads

$$(mc^2\sigma_3 - i\hbar c\sigma_1\partial_x)\Psi(x,t) = i\hbar\partial_t\Psi(x,t) \quad (48)$$

Where $\Psi(x,t)^T = (\psi_1(x,t), \psi_2(x,t))$ is a bicomponent spinor (their components are chirality amplitudes, since there is no spin in 1D) and the Pauli matrices were defined previously after equation (21).

We use the polar form of the spinor components as in the Pauli equation, so that each spinor component is expressed in polar form by the equation (24). The MdBB system of PDE equations, then, are

$$\frac{\partial R_1(x,t)}{\partial t} - c\sin\chi \frac{\partial R_2(x,t)}{\partial x} + \frac{c}{\hbar}\cos\chi R_2(x,t)\frac{\partial S_2(x,t)}{\partial x} = 0$$
$$\frac{\partial R_2(x,t)}{\partial t} - c\sin\chi \frac{\partial R_1(x,t)}{\partial x} - \frac{c}{\hbar}\cos\chi R_1(x,t)\frac{\partial S_1(x,t)}{\partial x} = 0 \quad (49)$$

$$\frac{\partial S_1(x,t)}{\partial t} - c\sin\chi \frac{R_2(x,t)}{R_1(x,t)}\frac{\partial S_2(x,t)}{\partial x} - \frac{\hbar c\cos\chi}{R_1(x,t)}\frac{\partial R_2(x,t)}{\partial x} + mc^2 = 0$$
$$\frac{\partial S_2(x,t)}{\partial t} + c\sin\chi \frac{R_1(x,t)}{R_2(x,t)}\frac{\partial S_1(x,t)}{\partial x} - \frac{\hbar c\cos\chi}{R_2(x,t)}\frac{\partial R_1(x,t)}{\partial x} + mc^2 = 0 \quad (50)$$

Where we have defined the angle-like quantity $\chi = \chi(x,t) = \frac{S_1(x,t) - S_2(x,t)}{\hbar} - \frac{\pi}{2}$.

Equations (49) and (50) are a set of four equations, two for the evolution of the modulus of the spinors and two for the evolution of their phases. Due to the particular expression of Weyl's equation, in the MdBB equations (49)-(50), it happens that the functions of one spinor depend mostly in the functions of the other spinor, the only dependency with itself being through the angle $\chi(x,t)$. Consequently, we can't treat each equation by its own, but treat as belonging to a system of two equations, (49) or (50).

The characteristic curves are then given by the eigenvalues of the matrices

$$\frac{\partial G_i}{\partial R_j} = \begin{pmatrix} 0 & -c\sin\chi \\ -c\sin\chi & 0 \end{pmatrix}$$

$$\frac{\partial G_i}{\partial S_j} = \begin{pmatrix} 0 & -c\frac{R_2}{R_1}\sin\chi \\ c\frac{R_1}{R_2}\sin\chi & 0 \end{pmatrix} \quad (51)$$

Where $i, j = 1,2$. Therefore,

$$\frac{dx_1}{dt} = c \sin \chi$$
$$\frac{dx_2}{dt} = -c \sin \chi \tag{52}$$

The guiding equations (52) constitute a new proposal up to our knowledge. It presents two interesting features. First, the spinor components have opposite velocities, showing a well-known feature of this equation. Second, the velocity does not depend on the gradient of actions, even more, it appear to be constant. The reason is that Weyl equation only includes first-order derivatives with respect to positions, and the dependence of the velocity with the gradient of the action is, in fact, a consequence of the second derivatives with respect to position. Even more, guiding equations (52) show that the velocity of the trajectories is related with the phase difference of the wave functions in the position, so that the velocity goes from zero if $S_1 - S_2 = (2n+1)h/4$ to $v = \pm c$ if $S_1 - S_2 = nh/2$ ($n$ being an integer).

Finally, equations (49)-(50) can be combined in their characteristic form,

$$\frac{d[R_1(x,t) \mp R_2(x,t)]}{dt} = \pm \frac{c}{\hbar} \cos \chi \left[ \frac{\partial S_1(x,t)}{\partial x} - \frac{\partial S_2(x,t)}{\partial x} \right]$$
$$\frac{dS_1(x,t)}{dt} \mp \frac{R_2(x,t)}{R_1(x,t)} \frac{dS_2(x,t)}{dt} = \pm \frac{\hbar c}{R_1(x,t)} \left[ \frac{\partial R_1(x,t)}{\partial x} + \frac{\partial R_2(x,t)}{\partial x} \right] + \left[ 1 \mp \frac{R_2(x,t)}{R_1(x,t)} \right] mc^2 \tag{53}$$

Where both equations (53) are valid along the trajectories $\frac{dx}{dt} = \pm c \sin \chi$.

Before continuing, there is an alternative that worth to be investigated. MdBB equations (49)-(50) are first order partial differential equations with respect to all the variables. It is possible, then, to apply the theory of characteristic to the whole ensemble. The characteristic curves are then the eigenvalues of the matrix

$$\begin{pmatrix} \frac{\partial G_i}{\partial R_i} & \frac{\partial G_i}{\partial S_i} \\ \frac{\partial G_j}{\partial R_i} & \frac{\partial G_j}{\partial R_i} \end{pmatrix} = c \begin{pmatrix} 0 & -\sin \chi & 0 & \cos \chi R_2/\hbar \\ -\sin \chi & 0 & -\cos \chi R_1/\hbar & 0 \\ 0 & -\frac{\hbar \cos \chi}{R_1} & 0 & -\sin \chi \frac{R_2}{R_1} \\ \frac{\hbar \cos \chi}{R_2} & 0 & -\sin \chi \frac{R_1}{R_2} & 0 \end{pmatrix} \tag{54}$$

Where $i = 1,2$ and $j = 3,4$. Therefore,

$$\frac{d x_1}{d t} = c$$
$$\frac{d x_2}{d t} = -c$$
(55)

With the characteristic form,

$$\pm \frac{\hbar}{R_1(x,t)\cos \chi}\left[\sin \chi \frac{d R_1(x,t)}{d t} + \frac{d R_2(x,t)}{d t}\right] + \frac{d S_1(x,t)}{d t} + mc^2 = 0$$
$$\mp \frac{\hbar}{R_2(x,t)\cos \chi}\left[\frac{d R_1(x,t)}{d t} - \sin \chi \frac{d R_2(x,t)}{d t}\right] + \frac{d S_2(x,t)}{d t} + mc^2 = 0$$
(56)

Where equations (56) are valid along the trajectories $\frac{d x}{d t} = \pm c$.

The guiding equation (55) was previously proposed by Gaveau et al.[29] using a hydrodynamic analogy. We think that further study is needed in order to see the advantages and disadvantages of both possibilities.

To finish, let us recall here a recent result by Holland.[28] For that, Weyl's equation (48) must be rewrite in another terms. Holland, using a technique described in reference 10,[30] adds three internal, Euler-like angles, to the spinor, so it can be described with positions, done by the vector **x**, and the Euler angles $\phi$ (a rotation), $\chi$ (azimuthal) and $\theta$ (polar angle). Although these angles can be related to the spinor components –in fact, the azimuthal angle $\chi$ is the angle we have already defined; they must be treated as new coordinates. Then Weyl's equation can be written as,[28]

$$i\hbar \partial_t \psi(t, x, \alpha) = -2c\hbar \lambda_i \partial_x \psi(t, x, \alpha)$$
(57)

Where $\lambda_i$ is a momentum operator that carries the derivatives with respect to the angles and $\alpha = (\phi, \chi, \theta)$ is a shortcut to the angle vector.

Writing $\psi(t, x, \alpha)$ in polar form we arrive to the corresponding MdBB equations,

$$\partial_t R + \frac{2c}{\hbar}(\lambda_i R \partial_x S + R \lambda_i \partial_x S + \partial_x R \lambda_i S) = 0$$
(58)

$$\partial_t S - \frac{2c}{\hbar}\left(\hbar^2 \frac{\lambda_i \partial_x R}{R} - \partial_x S \lambda_i S\right) = 0$$
(59)

Application of equation (8) gives in this case,

$$\frac{d\mathbf{x}}{dt} = \frac{\partial G_1}{\nabla_{\partial_i R} R} + \frac{\partial G_1}{\nabla_{\lambda_i R} R} = \frac{\partial G_2}{\nabla_{\partial_i S} S} + \frac{\partial G_2}{\nabla_{\lambda_i S} S} = \frac{2c}{\hbar}\left(\partial_x S + \lambda_i S\right) \tag{60}$$

Equation (60) shows the guiding equation as having two components. Since we have used to set of variables, i.e., positions and angles, we can see these two components as linear an angular velocities (linear velocities given by the derivative with respect to the angle coordinates and vice versa). Of course, equation (60) is the guiding equation obtained recently by Holland.[28] The fact that we have obtained them with the method of characteristics, in spite of the addition of the new angular coordinates, reinforced the idea that the guiding equation is the result of applying the method of characteristics to the MdBB system of PDE equations.

## IV. A note on the existence of the complete integral and the total differentiability of the quantum action

In previous sections, we have proved that the guiding equation is a property of the MdBB system of PDE. In fact, it is an example of Generalized Cauchy problem, solved by theory of characteristic curves. In that way, we have seen that quantum trajectories are, in fact, the characteristic curves of MdBB equations. Although the existence of such trajectories has been demonstrated elsewhere,[31,32] we want to address in this last section two important questions of the Theory of Characteristics, namely the fact that dR and dS are total differentials and that the solutions can be built via total integrals.

These two points are important because they are the usual way to prove the existence of the characteristic curves themselves in the theory of characteristic. Effectively, following Carathéodory,[15] the existence of the characteristic curves relies in the fact that $dS$ and $dR$ would be total differentials; because then, the surfaces $S$ and $R$, the solutions, can be built as the envelopes to the tangent planes to the family of surfaces themselves parameterized by the characteristic curves.

We will prove that $dS$ and $dR$ are total differentials by two ways. First, with a direct derivation from the S-equation (19) and, secondly, by proving that there is a hamiltonian $H(R, S, \nabla R, \nabla S)$ from which MdBB equations are the hamiltonian equations in action-angle variables. The first idea allows also finding how the tangent to the surface varies, so gaining insight in its evolution. The second allows a new derivation of MdBB equations, and show that both equations have the same origin and can be understood as veritable Hamilton-Jacobi equations. We will limit ourselves to the Schrödinger case, since the others can be done analogously.

*IV.1. Total differentials through direct calculation*

The initial point of the direct calculation is the S-equation (19), which we reproduced here,

$$\frac{\partial S(\mathbf{x},t)}{\partial t} + \frac{[\nabla S(\mathbf{x},t)]^2}{2m} - \frac{\hbar^2}{2m}\left\{\frac{1}{4m^2}\left\{\int_{t_0}^{t}\nabla[\nabla^2 S(\mathbf{x}(t'),t')]dt'\right\}^2 - \right.$$
$$\left. -\frac{1}{2m}\int_{t_0}^{t}\nabla^4 S(\mathbf{x}(t'),t')dt'\right\} + V(\mathbf{x},t) = 0 \qquad (19)$$

Equation (19) is equivalent to the MdBB equations (1)-(2). It is also a PDE for the action, so that we can apply the theory of characteristic to it. Application of equations (7) to equation (19) says that, if it is parameterized with the time, it is equivalent to the system of ODE,

$$\frac{d\mathbf{x}}{dt} = \frac{\nabla S(\mathbf{x},t)}{m}$$
$$\frac{d[\nabla S(\mathbf{x},t)]}{dt} = -\nabla V(\mathbf{x},t) - \frac{\hbar^2}{2m}\nabla\left\{\frac{1}{4m^2}\left\{\int_{t_0}^{t}\nabla[\nabla^2 S(\mathbf{x}(t'),t')]dt'\right\}^2 - \frac{1}{2m}\int_{t_0}^{t}\nabla^4 S(\mathbf{x}(t'),t')dt'\right\}$$
$$\frac{dS(\mathbf{x},t)}{dt} = \nabla S\frac{d\mathbf{x}}{dt} + \frac{\partial S(\mathbf{x},t)}{\partial t}$$

(61)

The first of equations (61) is the guiding equation. The second of equations (61) is the same that equation (14) but expressed only with actions. It is important to note that, since the biharmonic operator appears in (61), we are forced to demand continuity up to the fourth derivative for the action. Even more, because the quantum potential involves the laplacian of the modulus of the wave function, we must require also continuity up to the sixth derivative of the modulus of the wave function.[33]

The second equation of (61) also shows the special behavior of Quantum Physics, since it is the only equation formally different to its classical counterpart. It's worth highlighting, first, that all the evolution up to an instant is needed to compute the future evolution (this is the responsible of the why computations in Quantum Physics are so harder than in Classical Physics, where the trajectories can be computed independently). Second, it adds effects due to derivatives higher than the second order, which are difficult to foresee, but which, are, at the end, responsible for the quantum behavior. Finally, the theory doesn't show clearly the evolution of the derivatives of higher order, which is, in fact, a consequence that the non-locality of the equations.

The third of the equations (61) is the searched result. It shows that $dS$ is a total differential, and then that the characteristic curves exist.

This also means that the evolution of all quantities in the MdBB system of PDE equations can be computed. In section I we have computed the evolution of $R(\mathbf{x},t)$ and $S(\mathbf{x},t)$, with there first derivatives. It rests the second derivatives. From the Hamilton-Jacobi equation we have,

$$\frac{\partial S(\mathbf{x},t)}{\partial t} + \frac{[\nabla S(\mathbf{x},t)]^2}{2m} + V(\mathbf{x},t) - \frac{\hbar^2}{2m}\frac{\nabla^2 R(\mathbf{x},t)}{R(\mathbf{x},t)} = 0 \qquad (62)$$

We rewrite it as,

$$R(\mathbf{x},t)\Sigma(\mathbf{x},t) - \nabla^2 R(\mathbf{x},t) = 0 \tag{63}$$

Total derivative with respect to time gives

$$\frac{dR(\mathbf{x},t)}{dt}\Sigma(\mathbf{x},t) + \frac{d\Sigma(\mathbf{x},t)}{dt}R(\mathbf{x},t) = \frac{d[\nabla^2 R(\mathbf{x},t)]}{dt} \tag{64}$$

So that, if $R(\mathbf{x},t) \neq 0$ and $\Sigma(\mathbf{x},t) \neq 0$, and we use equation (63), equation (64) can be written as

$$\frac{1}{R(\mathbf{x},t)}\frac{dR(\mathbf{x},t)}{dt} + \frac{1}{\Sigma(\mathbf{x},t)}\frac{d\Sigma(\mathbf{x},t)}{dt} = \frac{1}{\nabla^2 R(\mathbf{x},t)}\frac{d[\nabla^2 R(\mathbf{x},t)]}{dt} \tag{65}$$

Equation (65) can be easily solved to obtain the evolution of the second derivative, as desired,

$$\nabla^2 R(\mathbf{x},t) = \frac{\nabla^2 R(\mathbf{x},0)}{R(\mathbf{x},0)\Sigma(\mathbf{x},0)} R(\mathbf{x},t)\Sigma(\mathbf{x},t) \tag{66}$$

Analogously, using the continuity equation (1) the evolution of the laplacian of the action is,

$$\nabla^2 S(\mathbf{x},t) = \frac{R(\mathbf{x},0)^2 \nabla^2 S(\mathbf{x},0)}{r(\mathbf{x},0)} \frac{r(\mathbf{x},t)}{R(\mathbf{x},t)^2} \tag{67}$$

With $r(\mathbf{x},t) = 2m[\partial R(\mathbf{x},t)/\partial t + \nabla R(\mathbf{x},t)\cdot \nabla S(\mathbf{x},t)/m]$. Equations (66)-(67), together with the evolution of the functions themselves, equations (18) and (20), and the evolution of their gradients, equation (16) and the analog equation for $\nabla R(\mathbf{x},t)$, are the complete set of ODE that solves the generalized Cauchy problem of the MdBB system of PDE.

*IV.2. Complete integrals through Hamilton-Jacob theory*

The problem of the complete integral of MdBB equations can be related to the Hamilton-Jacobi theory of PDE. The point of depart is the lagrangian density of Schrödinger's equation,

$$\mathsf{L}(\Psi,\Psi^*) = \frac{\hbar^2}{2m}\nabla\Psi\cdot\nabla\Psi^* - \frac{\hbar}{2i}\left(\frac{\partial \Psi^*}{\partial t}\Psi + \frac{\partial \Psi}{\partial t}\Psi^*\right) - \Psi^* V \Psi \tag{68}$$

Substitution of the wave function in the polar form, and integration by parts, make able to change the lagrangian density (68) by its equivalent,

$$\mathsf{L}\left(R, S, \nabla R, \nabla S, \frac{\partial R}{\partial t}, \frac{\partial S}{\partial t}\right) = \frac{\hbar^2}{m} R \nabla^2 R - \frac{R^2 (\nabla S)^2}{2m} + i\hbar R \frac{\partial R}{\partial t} - R^2 \frac{\partial S}{\partial t} - R^2 V \qquad (69)$$

In equations (68) and (69), and for now on, the explicit dependencies of the functions in **x** and *t* has not been shown for shake of simplicity in the expressions.

Direct computation can show that the Euler-Lagrange equations derived from the lagrangian density (69) are the MdBB equations. Alternatively, the momentum densities are,

$$\Pi_R \equiv \frac{\partial \mathsf{L}\left(R, S, \frac{\partial R}{\partial t}, \frac{\partial S}{\partial t}, \nabla R, \nabla S\right)}{\partial(\partial R / \partial t)} = i\hbar R$$

$$\Pi_S \equiv \frac{\partial \mathsf{L}\left(R, S, \frac{\partial R}{\partial t}, \frac{\partial S}{\partial t}, \nabla R, \nabla S\right)}{\partial(\partial S / \partial t)} = -R^2 \qquad (70)$$

The hamiltonian density is then computed by the Legendre transformation, provided that the Legendre transformation can be applied to this case, which in turn is the problem if an envelope of surfaces exist,

$$\mathsf{H}\left(R, S, \nabla R, \nabla S, \frac{\partial R}{\partial t}, \frac{\partial S}{\partial t}\right) = -\frac{\hbar^2}{m} R \nabla^2 R - \frac{R^2 (\nabla S)^2}{2m} + R^2 V + i\hbar \frac{R}{m} \nabla R \cdot \nabla S \qquad (71)$$

The hamiltonian density (71) was first introduced by Takabayasi;[34] it allowed him to obtain just one of the MdBB equation, the Hamilton-Jacobi equation. In order to obtain both equations, it suffices with remembering that we can add to the hamiltonian density (72) whatever term whose contribution to the hamiltonian was zero. Of course, it is the case of the continuity equation; hence, before substituting the densities of momentum (70) in the density hamiltonian (71) we add to it the term,

$$-\frac{\nabla R \cdot \nabla S}{m} - \frac{R \nabla^2 S}{m}$$

Then, the hamiltonian density can be expressed as,

$$\mathsf{H}(R, S, \nabla R, \nabla S, \Pi_R, \Pi_S) = \left(\frac{\hbar^2}{2m} \frac{\nabla^2 R}{R} + \frac{(\nabla S)^2}{2m} - V\right) \Pi_S + \left(\frac{\nabla R \cdot \nabla S}{m} - \frac{R \nabla^2 S}{m}\right) \Pi_R -$$
$$-\frac{\hbar^2}{2m} R \nabla^2 R - i\hbar R \frac{\nabla R \cdot \nabla S}{m} \qquad (72)$$

From the density hamiltonian (72), the Hamilton equations of motion are the four equations (now writing explicitly all the dependencies),

$$\frac{\partial [R(\mathbf{x},t)^2]}{\partial t} + \nabla \cdot \left[ \frac{R(\mathbf{x},t)^2 \nabla S(\mathbf{x},t)}{m} \right] = 0 \tag{73a}$$

$$\frac{\partial S(\mathbf{x},t)}{\partial t} + \frac{[\nabla S(\mathbf{x},t)]^2}{2m} + V(\mathbf{x},t) - \frac{\hbar^2}{2m} \frac{\nabla^2 R(\mathbf{x},t)}{R(\mathbf{x},t)} = 0 \tag{73b}$$

$$\frac{d \Pi_R}{dt} = \Pi_R \left( -\frac{\nabla^2 S}{m} \right) \tag{73c}$$

$$\frac{d \Pi_S}{dt} = 0 \tag{73d}$$

Substitution of the momentum densities definitions (70) shows that equation (73c) is the same as equation (73a) and that equation (73d) is the conservation of density, i.e., Liouville's theorem.

In this way, we have related the use of characteristic curves in MdBB equations to the Hamilton-Jacobi theory of PDE. Consequently, if a specific solution $u(\mathbf{x},t,\mathbf{a})$ is known, where $u$ stands for $R$ or $S$, and $\mathbf{a} = (p_x, p_y, p_z)$ are parameters that will be identified with the momenta, i.e., if we have a complete integral of the problem, then, the envelope of an arbitrary three parameters family of these solutions is also a solution. In fact, introducing the new parameters $\mathbf{b} = \nabla_\mathbf{a} u$, one obtains the six parameters family of solutions of the Hamilton equations of motion.

Finally, we want to note that some differences usually pointed out between both MdBB equations (1)-(2) are not so different. First, both have the same origin, as the evolution of the "coordinated" part of Hamilton equations. Second, although normally the equation (2) is named "Hamilton-Jacobi-like" equation, from the hamiltonian (72) is easy to see that both equations (1) and (2) are Hamilton-Jacobi equations. The extra term does not change the type of the equation (2), which continuous to be a Hamilton-Jacobi equation.[35] Third, both equations (1)-(2) are cyclic. The definition of the momenta (70) shows that in both cases, the conserved magnitude is the density $R(\mathbf{x},t)$, ad that it is the angle-like variable conjugated to the action.

### V. Conclusions

In the preceding sections, we have solved the generalized Cauchy problem stated by the MdBB system of PDE by means of the theory of characteristics. As a result, a system of ODE equations have appear, which is equivalent to the original system of PDE. Although solving the system of ODE has no reason to be easier than solving the original system of PDE, what is important is the form of some of the ODE, which are, indeed, properties of the original MdBB system of PDE.

In particular, the guiding equation appears in the form of equation (8) as the direction of the ray of characteristic curves, making natural the identification of the characteristic curves, a mathematical property of the MdBB system of PDE, with quantum trajectories, a physical property of the MdBB equations understood as the matter field equations of the problem under study. In most cases, also, the trajectories can be obtain also from a generalization of Newton's second law, which appears like one of the ODE equivalent to the MdBB system of PDE.

Moreover, the method has proved to be able to treat other evolution equations. In fact, since it application is straight forward once the corresponding MdBB system of PDE has been obtained, the methodology here presented is able to be generalized to whatever quantum evolution equation.

Finally, the application of the characteristic theory to the MdBB equations helps also to identify the origin of key properties of the theory. In particular, the non-locality appears naturally, since we always need derivatives of higher order to solve completely the problem, as shown in section IV. Also, the usual relation between the velocity field and the gradient of the action has revealed to be a consequence of the existence of second derivatives in the evolution equation, whereas it should be related directly with the action when there are only gradients.

**Acknowledgments**

X.G. and J.M.B. acknowledge financial support from the Spanish *Ministerio de Ciencia y Tecnología*, DGI project CTQ2005-01117/BQU and, in part from the *Generalitat de Catalunya* projects 2005SGR-00111 and 2005SGR-00175.